\documentclass[showpacs,preprintnumbers,twocolumn,amsmath,amssymb]{revtex4}
\usepackage{dcolumn}
\usepackage{bm}
\usepackage{epsfig,graphicx}
\usepackage{times}

\begin{document}

\title{Description of dynamics of stock prices by a Langevin approach}

\author{Zi-Gang Huang{$^1$}, Yong Chen{$^{1}$}, Yong Zhang{$^2$}, and Ying-Hai Wang{$^{1}$}}

\address {$^1$Institute of Theoretical Physics, Lanzhou University, Lanzhou $730000$, China\\
$^2$Department of Physics, Center for Nonlinear Studies, and The Beijing-Hong Kong-Singapore Joint Center for Nonlinear and Complex Systems (Hong Kong), Hong Kong Baptist University, Kowloon Tong, Hong Kong, China}

\date{\today}

\begin{abstract}
We present a time-dependent Langevin description of dynamics of
stock prices. Based on a simple sliding-window algorithm, the
fluctuation of stock prices is discussed in the view of a
time-dependent linear restoring force which is the linear
approximation of the drift parameter in Langevin equation
estimated from the financial time series. By choosing suitable
weighted factor for the linear approximation, the relation between
the dynamical effect of restoring force and the autocorrelation of
the financial time series is deduced. We especially analyze the
daily log-returns of S$\&$P $500$ index from $1950$ to $1999$. The
significance of the restoring force towards the prices evolution
are investigated from its two coefficients, slope coefficient and
equilibrium position. The new simple form of the restoring force
obtained both from statistical and theoretical analyses suggests
that the Langevin approach can effectively present the
macroscopical and the detail properties of the price evolution.
\end{abstract}

\maketitle

\section{\label{sec:level1} Introduction}

The analysis of financial data by methods developed for physical
systems has a long tradition \cite{Bachelier, Pareto, Levy,
Mandelbrot}, and has attracted the interest of physicists. One of
the most motivated reasons is that it is a great scientific
challenge to understand the dynamics of a strongly fluctuating
complex system with a large number of interacting elements. In
addition, it may be possible that the experience gained in the
process of studying complex physical systems might yield new
results in economics.

There are many observables generated from financial markets, and
one central issue of the research on the dynamics of financial
markets is the statistics of price changes which determine losses
and gains. The price changes of a time series of quotations $x(t)$
are commonly measured by returns: $r:={x(t+\tau)}/{x(t)}$,
log-returns, or increments: $\Delta{x}:=x(t+\tau)-x(t)$ at a time
scale $\tau$. In 1900, Bachelier proposed the first model for the
stochastic process of returns --- an uncorrelated random walk with
independent identically distributed Gaussian random variables
\cite{Bachelier}. However, prices do not follow a signal random
walk process \cite{Matengna, Zhang, Liu, Gopikrishnan}. For
example, the daily correlation has been known as the daily
log-returns correlated with themselves in such a way that positive
returns are followed by positive returns as well \cite{Fama,
LeBaron}. Many considerations have been aroused by this effect
recently and meanwhile the related research has been reported, not
only for daily data \cite{Boguna} but also for high frequency data
\cite{Ohira, Sazuka}.

The Langevin equation (LE) which distinguishes the development of
sample path into the deterministic and random terms, has been used
to deal with the Brownian motion problem. Recently, the Langevin
approach was used to analyze the financial time series on scale
\cite{Friedrich,Ivanova}. Friedlich \emph{et al.} \cite{Friedrich}
have investigated how price changes $\Delta{x}$ on different time
scales $\tau$ are correlated motivated by hierarchical structure
of financial time series, which is similar to the energy cascade
in hydrodynamic turbulence. They derived a multiplicative Langevin
equation from a Fokker-Planck equation (FPE) in the variable scale
$\tau$ and performed the statistical way to distinguish and
quantify the deterministic and the random influence on the
hierarchical structure of the financial time series in terms of
the drift and diffusion parameters, $D^{(1)}$ and $D^{(2)}$,
respectively \cite{Friedrich}.

Different from the former study in which the LE is used to analyze
the scales evolutions of finance, yet, in this paper, with the
Langevin description, a new insight on the dynamics of the process
will be obtained by investigating the time-dependence of
log-returns. The time-dependent properties of prices evolution are
derived in the way of estimating drift parameter $A(z)$ of sampled
local periods in the sliding window. Then, the relation between
$A(z)$ and autocorrelation $C$, average return
$\langle{z}\rangle$, from which the practical significance of
$A(z)$ can be recognized, are resulted both from the statistical
time-dependence of $A(z)$ and some theoretical analyses. Besides,
our Langevin description contains, as a particular case with
flat-$A(z)$, the effect of daily correlation in log-returns. On
the other hand, the form of diffusion parameter $B(z)$ got in this
paper, to some extent, explains the heavy tailed probability
densities of price changes.

The research are mainly carried out from the samples of the daily
log-returns $z(t)|_{\tau=1d}$ of S$\&$P $500$ index from $1950$ to
$1999$, containing $12583$ days, thus covering a wide time range
with many different economic and political situations.

The paper is organized as follows. In Sec. II, taking the daily
log-returns as an example, we generally discuss the application of
Langevin approach to log-returns series. In Sec. III, we show the
results and discussions. Finally, the summary and the outlook of
this paper are given in Sec. IV.

\section{\label{sec:level2} The Langevin approach to log-returns series}

For a time series of prices or market index values $x(t)$, the
log-return $z(t)\equiv{z(t)_{\tau}}$ over a time scale $\tau$ is
defined as the forward change in the logarithm of $x(t)$,
\begin{equation}
z(t)_{\tau}\equiv{lnx(t+\tau)-lnx(t)}.\label{eq00}
\end{equation}
The behavior of daily log-return $z(t)_{1d}$ as a stochastic
variable is described by the following LE:
\begin{equation}
dz=A(z)dt+B(z)dw,\label{eq01}
\end{equation}
where the drift parameter $A(z)$ and diffusion parameter $B(z)$
respectively describe the deterministic and the random influences
on the time process of log-returns, and $dw$ denotes the increment
of a standard Wiener process. It is assumed that, within each
sampling window the parameters may depend on the log-returns, but
not explicitly on time (stationary). Thus, the drift and diffusion
parameters of the sampled period can be extracted from the sampled
data by simply using the definition \cite{Sura},
\begin{eqnarray}
A(z)&=&\lim_{\Delta{t}\rightarrow
0}\frac{1}{\Delta{t}}\langle{Z(t+\Delta{t})-z}\rangle|_{Z(t)=z}\label{eq02}\\
B(z)^{2}&=&\lim_{\Delta{t}\rightarrow
0}\frac{1}{\Delta{t}}\langle{(Z(t+\Delta{t})-z)^{2}}\rangle|_{Z(t)=z} \label{eq03}.
\end{eqnarray}
Here $\langle{\cdots}\rangle$ denotes the averaging operator and
$Z(t+\Delta{t})$ is a realization of the LE (\ref{eq01}). From
Eq.(\ref{eq02}), it is obvious that the drift parameter $A(z)$ is
the average increment of unit time under the condition $Z(t)=z$,
which represents the deterministic influences. $B(z)$ is the
deviation of $A(z)$ which pictures the random influences. It has
been known that the autocorrelation of the log-returns decays very
fast which is usually characterized by a correlation time much
shorter than a trading day \cite{Matengna}. When the time
increment $\Delta{t}$ is larger than $1$ day, the daily
log-returns can be considered as the result of many uncorrelated
`shocks'. Thus, in this paper, $\Delta{t}$ is mainly set as $1$
day. Compared to the length of time window $T$, $\Delta{t}=1d$
approximately accords with the limit in Eq.(\ref{eq02}) and
Eq.(\ref{eq03}).

\begin{figure}
\centerline{\resizebox{9cm}{!}{\includegraphics{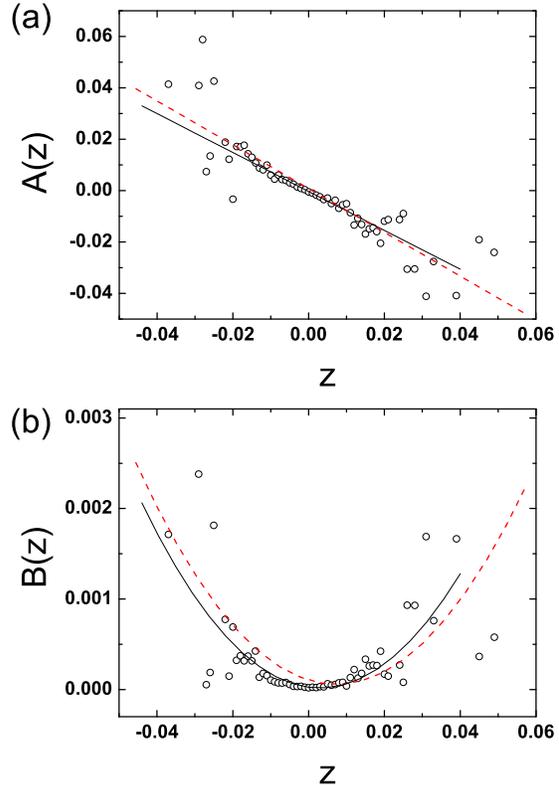}}}
\caption{(Color online) The statistical results of $A(z)$ and
$B(z)$ (circles) estimated from the daily log-returns of S$\&$P
$500$ index from $28$ May. $1962$ to $10$ Jun. $1970$, a window
with length $T=2000$ days. The approximations of the statistical
results with weight factor $P(z_{i})$ (solid lines) and that with
equal-weight (dashed lines) respectively correspond to the fitting
errors $\sigma _{A}=2.564\times{10^{-4}}$ and $8.650
\times{10^{-3}}$ and  $\sigma_{B}=1.336\times{10^{-5}}$ and
$5.144\times{10^{-4}}$. The expressions of solid lines are,
$A(z)=-0.7569z-3.24305\times{10^{-4}}$ and
$B(z)=0.92478z^{2}-0.00562z+2.48829\times{10^{-5}}$.} \label{fig1}
\end{figure}

Based on the samplings all over the long series, the statistical
results of $A(z)$ and $B(z)^{2}$ which are respectively estimated
by Eq. (\ref{eq02}) and Eq. (\ref{eq03}) have their simple and
general forms. The results of $A(z)$ are close to a linear form,
and that of $B(z)^{2}$ are close to a parabolic form,
\begin{eqnarray}
A(z)&=&az+b\label{eq04}\\
B(z)^{2}&=&a'z^{2}+b'z+c'.\label{eq05}
\end{eqnarray}
Fig. \ref{fig1} presents the statistical results of $A(z)$ and
$B(z)^{2}$ (circles) which are estimated from the daily
log-returns time series of S$\&$P $500$ index from $28$ May.
$1962$ to $10$ Jun. $1970$ (window length $T=2000d$). The
statistical results for large $z$ are more noisy and uncertain
than the points near the origin, because these border points are
visited rarely by the trajectory. On the contrary, as viewed from
statistics, the more one given $z_{i}$ is visited, the more times
the averaging operator $\langle{\cdots}\rangle|_{Z(t)=z_{i}}$ in
Eq. (\ref{eq02}) works, which would produce more accurate and
reasonable $A(z_{i})$. Thus, while approximating $A(z_{i})$
($i=1,2,...,n$) with linear form and $B(z_{i})^{2}$
($i=1,2,...,n$) with parabolic form by Least-squares fit, the
effect of the visited probability for each $z_{i}$ should be
considered, with each $z_{i}$ corresponding to its own weight.

It is natural for a physical scientist to define the weighted
factor of $z_{i}$ as its probability,
\begin{equation}
P(z_{i})=N(z_{i})/T,\label{eq06}
\end{equation}
where the $N(z_{i})$ is the frequency of $z_{i}$ within the time
window of length $T$. Fig. \ref{fig1} shows the approximations of
the statistical results of $A(z)$ and $B(z)^{2}$ with equal-weight
(dashed lines) and weighted factor $P(z_{i})$ (solid lines). The
values of the fitting error $\sigma_{A}$ and $\sigma_{B}$ (the
mean standard deviations from approximations of $A(z)$ and
$B(z)^{2}$) with weighted factor $P(z_{i})$ are visibly lower than
those with equal-weight. Therefore, while approximating the
statistical results, the point at $z_{i}$ is endued the weight
$P(z_{i})$ which is correlated to the frequency of $z_{i}$ within
the window.

\begin{figure}
\centerline{\resizebox{10cm}{!}{\includegraphics{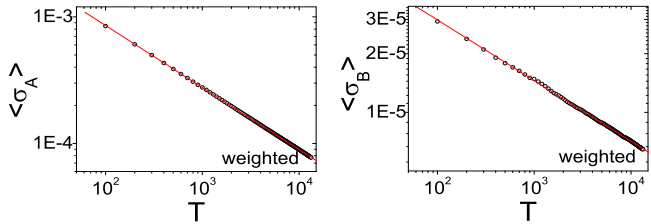}}}
\caption{(Color online) Log-log plots of weighted fitting errors
$\sigma_{A}$ and $\sigma_{B}$ vs. $T$, with $T=100d\sim 13000d$.
They exhibit perfect power-law behavior. The solid lines are
power-law regression fits over the entire range, which give
estimates of power-law exponents, $\eta_{A}=0.49022$ and
$\eta_{B}=0.31122$.} \label{fig2}
\end{figure}

For the given return value $z_{i}$, the frequency $N(z_{i})$ also
depends on the length of time windows. Thus, it is worth
mentioning the influences from the length $T$ on the
approximations. The relations between $T$ and the fitting errors
of $A(z)$ and $B(z)^{2}$ are investigated. To estimate the error
purely made by window length $T$, not by information the time
series embodied, the daily log-returns series is randomly
rearranged and the fitting errors $\sigma_{A}$ and $\sigma_{B}$
vs. $T=100d\sim13000d$ are calculated (Fig. \ref{fig2}). It was
found that the weighted fitting errors $\sigma_{A}$ and
$\sigma_{B}$ of the approximations with weighted factor $P(z_{i})$
decline quickly with a perfect power-law behavior,
$\sigma_{A}=8.178\times{10^{-3}}\cdot T^{-0.49022}$ and
$\sigma_{B}=1.249\times{10^{-4}}\cdot T^{-0.31122}$. This behavior
suggests that the statistical results of $A(z)$ (or $B(z)^{2}$)
with larger $T$ is more feasible to be approximated with linear
(or parabolic) form. Note incidentally that, only the results with
the same window length $T$ could be compared since different $T$
corresponds to different fitting errors.

The algorithm for the detection of the time-dependence of drift
term $A(z)$ in this paper can be described as follows: sample the
long log-returns series with a sliding window of short length T,
and compute $A(z)$ for each location. The results estimated from
the time window which samples a given local period, present the
corresponding local characters of financial markets. The algorithm
is more sensitive than merely studying transient behavior. The
comment for the selection of window length $T$ is to choose $T$
long enough so that the averages in Eq. (\ref{eq02}) and Eq.
(\ref{eq03}) are statistically meaningful but not so long as to
lose the temporal resolution. In the results presented below,
$250d$ and $2000d$ window lengths and $5d$ overlapping (window
shift by $5$ days per time) are used. The corresponding fitting
errors are, $\sigma_{A,250d}=5.46\times{10^{-4}}$,
$\sigma_{B,250d}=2.24\times{10^{-5}}$, and
$\sigma_{A,2000d}=1.97\times{10^{-4}}$,
$\sigma_{B,2000d}=1.17\times{10^{-5}}$. It is expected that the
variation of $A(z)$ as a function of time can accurately indicate
interesting dynamical changes in financial process.

\section{\label{sec:level3} Results and discussions}

\begin{figure}
\centerline{\resizebox{9cm}{!}{\includegraphics{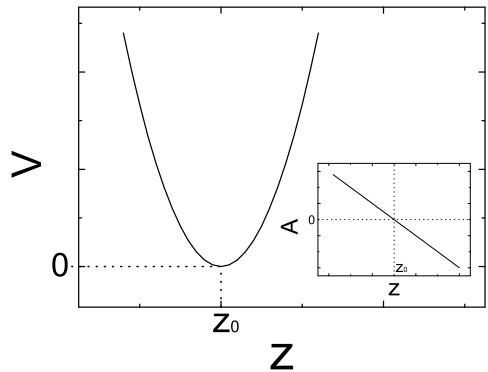}}} \caption{ The
sketch maps of the linear restoring force $A(z)$ and its
corresponding potential well $V(z)$.} \label{fig3}
\end{figure}

\begin{figure}
\centerline{\resizebox{9cm}{!}{\includegraphics{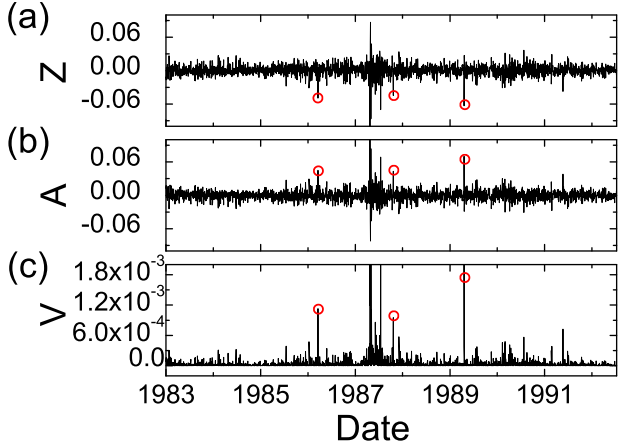}}}
\caption{(Color online) (a) Log-returns series $z(t)$, and (b) the
corresponding restoring forces $A$ , and (c) the potential $V$.
Restoring force and potential at one given time $t_{i}$ is defined
as $A(t_{i})\equiv{A(z(t_{i}))}$ and
$V(t_{i})\equiv{V(z(t_{i}))}$. The $A(z)$ and $V(z)$ are the
average results calculated from these $T=250d$ overlapping windows
so far as embodying the time $t_{i}$. Some special points are
marked with circles to reflect relations between the three
series.} \label{fig4}
\end{figure}

In the LE (\ref{eq01}), the drift parameter $A(z)$ could be seen
as an action of potential, with $A(z)=-\nabla{V(z)}$. Noted that
$A(z)$ has a linear form with negative slope, it could be
interpreted as the effect of linear state-dependent restoring
force with symmetrical potential well,
\begin{equation}
V(z)=-\frac{a}{2}z^{2}-bz-\frac{b^{2}}{2a} \label{eq09},
\end{equation}
where $z$ presents the position of one given particle enslaved to
it. The sketch maps of $A(z)$ and $V(z)$ are showed in Fig.
\ref{fig3} in which the equilibrium position ($A(z_{0})=0$) is
$z_0=-b/a$. Fig. \ref{fig4} shows the time series of log-returns
$z(t)$, restoring force $A(z)$ and potential $V(z)$ from $26$ May.
$1983$ to $1$ Jan. $1993$. It is easy to find that the vibrancy of
log-returns $z$ presented seems to be similar to force $A(z)$ and
potential $V(z)$, which can be clearly seen from several large
events marked in this figure, and the restoring force $A(z)$
always presents converse effect to log-returns $z$. In some large
events, the potential $V(z)$ is exceedingly large.

In the following, the time dependence of restoring force $A(z)$
will be discussed including the equilibrium position $z_0$ and the
slope coefficient $a$. In addition, discussions of the diffusion
parameter $B(z)$ and the error analysis will be given.

\subsection{\label{sec:level4} Equilibrium Position $z_0$}

\begin{figure}
\centerline{\resizebox{9cm}{!}{\includegraphics{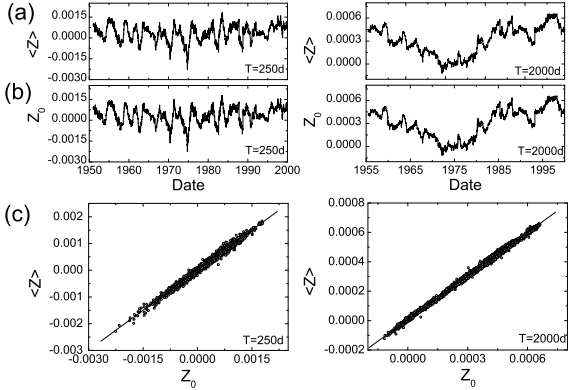}}}
\caption{Sliding window analysis of the time-dependence of (a) the
average log-return $\langle{z}\rangle$, and (b) the equilibrium
position $z_0$. The lengths of sliding windows are $T=250d$ (left)
and $T=2000d$ (right), and windows shift by $5$ days per time. (c)
The correlation function between $z_{0}$ and $\langle{z}\rangle$.
Their linear approximations (solid lines) are:
$\langle{z}\rangle=0.9985z_{0}+3.4060\times{10^{-6}}$ (left) with
standard deviation $SD=5.78\times{10^{-5}}$, and
$\langle{z}\rangle=0.9856z_{0}+7.8319\times{10^{-6}}$ (right) with
standard deviation $SD=1.00\times{10^{-5}}$.} \label{fig5}
\end{figure}

From Langevin equation \cite{Hanggi}, the so called equilibrium
position $z_0$, which is the zero value of negative-sloped linear
drift term, corresponds to the minimum of the potential well. From
a physical point of view, the average displacement
$\langle{z}\rangle$ of an oscillating particle in the potential
well defined by Eq. (\ref{eq09}) should also be the minimum of the
potential well, $(-b/a,0)$. Thus, we get,
\begin{eqnarray}
z_0 \simeq \langle{z}\rangle. \label{eq07}
\end{eqnarray}
The average displacement $\langle{z}\rangle$ was directly obtained
from the log-returns series, and the equilibrium position $z_0$
was calculated from Eq. (\ref{eq04}). The time dependence of $z_0$
and $\langle{z}\rangle$ coincide with each other very well all
over the ranges [see Fig. \ref{fig5}(a) and \ref{fig5}(b)], and
the plots of $z_0$ as a function of $\langle{z}\rangle$ in Fig.
\ref{fig5}(c) was excellent agreement with Eq. (\ref{eq07}).

In language of finance, the average log-return $\langle{z}\rangle$
describes the macroscopical trend of the price movement:
$\langle{z}\rangle>0$ indicates going up, and
$\langle{z}\rangle<0$ indicates going down. Therefore, based on
Eq. (\ref{eq07}), the rising trend of prices could be estimated if
the statistical result $z_0=-a/b>0$; otherwise, the falling trend
could be estimated if $z_0=-a/b<0$. Thus, the equilibrium position
$z_{0}$ of restoring force $A(z)$ is comprehended as the `trend
index' of stock prices. Furthermore, the stock prices and their
log-returns are macroeconomic indicators which are widely used
because of the strong correlation between financial markets and
economic development. In this case, the equilibrium position
$z_0$, which is derived from the LEs description of financial time
series, would be another important indicator of macroeconomics.

\subsection{\label{sec:level5} Slope Coefficient $a$}

\begin{figure}
\centerline{\resizebox{9cm}{!}{\includegraphics{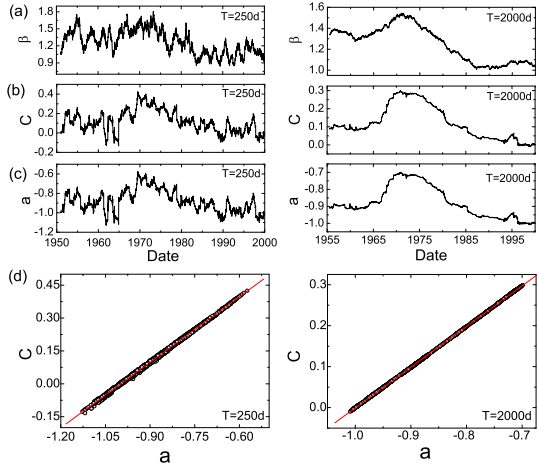}}}
\caption{(Color online) Sliding window analysis of the
time-dependence of (a) the proportion of sign-convert cases and
sign-sustained cases $\beta$, (b) the autocorrelation function
$C_{\Delta{t}=1}$, and (c) the slope coefficient $a$ with
$\Delta{t}=1d$. The lengths of sliding windows are $T=250d$ (left)
and $T=2000d$ (right), and windows shift by $5$ days per time. (d)
The correlation function between $a$ and $C$. Their linear
approximations (solid lines) respectively are: $C=0.9942+0.9958a$
(left) with standard deviation $SD=3.44\times{10^{-3}}$, and
$C=0.9932+0.9938a$ (right) with standard deviation
$SD=6.44\times{10^{-4}}$.} \label{fig6}
\end{figure}

In Fig. \ref{fig6}(c), the time dependence of the slope
coefficient $a$ calculated from the daily log-returns of S$\&$P
$500$ with $T=250d$ and $2000d$ are plotted. The ranges of the
slope coefficient $a$ with $T=250d$ and $2000d$ are
$[-1.127,-0.574]$ and $[-1.009,-0.699]$ respectively, both of
which are close to the value $-1$. As shown in Fig. (\ref{fig3})
we know that in a certain given position $z_{i}$, steeper
(flatter) slope of $A(z)$ corresponds to larger (smaller)
restoring force $A(z_{i})$. Thus, one can imagine the mechanism of
our model: it would take few times for larger forces (slope:
$a<-1$) to draw particles from one side of the equilibrium
position $z_{0}$ to another side, which we called
`$z_{0}$-crossing' action for the moment, and more times for
smaller forces (slope: $-1<a<0$). To discuss the aforementioned
mechanism more accurately, normalized daily log-returns are used,
\begin{equation}
g=\frac{z-\langle{z}\rangle_{T}}{\nu},
\nu=\sqrt{\langle{z^{2}}\rangle_{T}-\langle{z}\rangle^2_{T}}
\label{eq15},
\end{equation}
which has zero mean value, $\langle{g}\rangle_{T}=0$. Here the
standard deviation $\nu$ of log-returns is defined as the time
averaged volatility \cite{Gopikrishnan} and the
$\langle{\cdots}\rangle_{T}$ denotes an average over the entire
length of the series within time window $T$. From Eq.
(\ref{eq07}), $g$ has the equilibrium position $z_{0}$ equaling
zero, so that the $z_{0}$-crossing action could be reduced to the
sign convert of $g$. Thus the mechanism can be described as
follows: The sign of $g$ changes frequently while the slope is
quite steep; on the contrary, same signs congregate together and
sequences of consecutive `$+$' or `$-$' appear when the slope is
flat.

The sign series of daily log-returns has been considered to study
the daily correlation in log-returns \cite{Boguna}. Those
researchers got the conditional dynamics from the sequence of
consecutive `$+$' and `$-$'. In this paper, however, we will
compare the sequence of the same sign with the converting sign of
two neighboring days to check the relationship between $a$ and
$z_{0}$-crossing.

The sign-cases of a given day and its previous day are: `$++$',
`$+-$', `$-+$' and `$--$'. One can define `$++$' and `$--$' as the
sign-sustained cases, and `$+-$' and `$-+$' as the sign-convert
cases. Incidentally, the contribution of `$++$' to the
sign-sustained cases was shown to be a little more than `$--$' on
average over the whole series. Then, the time series of signs of
$g$ is investigated by counting the frequencies of sign-sustained
cases, $m$, and sign-convert cases, $n$. The time dependence of
the proportion $\beta=m/n$ and the slope coefficient $a$
calculated from the daily log-returns of S$\&$P $500$ with
$T=250d$ and $2000d$ are plotted in Figs. \ref{fig6}(a) and (c).
The good similarity of $\beta$ and $a$ proves that the slope
coefficient is related to the $z_{0}$-crossing action and reflects
the correlation of neighboring daily log-returns. In detail, in
one sampled period, while the slope of the restoring force is
flat, the given day's sign of log-return is more likely to be the
same as its previous day; on the country, while the slope is
steep, the given day's sign is more likely to be different from
its previous day. Thus, the mechanism of the daily correlation in
log-returns is qualitatively explained by the restoring force.

It's easy to notice from the time dependence of $a$ and $\beta$
[Fig. \ref{fig6}(a)(c)] over the whole series, most of the periods
have their slope $a$ larger than $-1$, and $\beta$ larger than
$1$, with mean values $\bar{a}|_{T=250d}=-0.883$,
$\bar{a}|_{T=2000d}=-0.874$, and $\bar{\beta}|_{T=250d}=1.242$,
$\bar{\beta}|_{T=2000d}=1.223$. These imply that, in practice, the
flat-slope restoring force is more prevalent than the steep-slope
one, and the case of sign-sustained is more than that of the
sign-convert. Consequently, from the general appearance of
sign-sustained cases, the same conclusion was reached as that of
\cite{Boguna}: the return of the price during a given day can be
correlated with the previous day, in particular with the sign of
the previous day. However, it is worth noticing that, the analysis
with Langevin approach is more general because it contains the
positive daily correlation as the particular case with flat-slope
restoring force, and non-correlation with steep-slope restoring
force.

On the other hand, the autocorrelation function $C(\Delta{t'})$, a
typically important statistics of stochastic processes, is always
used to investigate pairwise correlation of the log-returns of a
financial asset. In the following, we compared it with the slope
coefficient $a$ from mathematical relations and statistical
results. It is known that,
\begin{eqnarray}
C(\Delta{t'})
=\frac{\langle{z(t)\cdot{z(t+\Delta{t'})}}\rangle_{T}-\langle{z(t)}\rangle^{2}_{T}}
{\langle{z(t)^{2}}\rangle_{T}-\langle{z(t)}\rangle^{2}_{T}}, \label{eq08}
\end{eqnarray}
where $\langle{\cdots}\rangle_{T}$ denotes time averaging over all
the trading days within the sampled local period with length $T$,
$\Delta{t'}$ is time increment.

For the weighted linear Least-squares fit, which has been used to
approximate the statistical results of $A(z)$, the weighted
objective function \cite{Neter,Seber} can be written as,
\begin{equation}
q=\sum_{i=1}^{n}{W_{i}[y_{i}-(ax_{i}+b)]^{2}}. \label{eq14}
\end{equation}
where $W_{i}$ presents the weight of the point $(x_{i},y_{i})$.
The values of $a$ and $b$ corresponding to the minimal values of
function $q$ are something to be sought for. Thus from Eq.
(\ref{eq14}), the solution of slope coefficient $a$ in $y(x)=ax+b$
is achieved,
\begin{eqnarray}
a=\frac{\sum_{i}{W_{i}}\sum_{i}{x_{i}y_{i}W_{i}}-\sum_{i}{x_{i}W_{i}}\sum_{i}{y_{i}W_{i}}}
{\sum_{i}{W_{i}}\sum_{i}{x_{i}^{2}W_{i}}-(\sum_{i}{x_{i}W_{i}})^2}
\end{eqnarray}
In the previous discussion, the weight of $z_{i}$ was defined as
its probability $P(z_{i})$ (Eq. (\ref{eq06})). Thus one get,
\begin{widetext}
\begin{eqnarray}
a=\frac{\sum_{i}{P(z_{i})}\sum_{i}{z_{i}A(z_{i})P(z_{i})}-\sum_{i}{z_{i}P(z_{i})}\sum_{i}{A(z_{i})P(z_{i})}}
{\sum_{i}{P(z_{i})}\sum_{i}{z_{i}^{2}P(z_{i})}-(\sum_{i}{z_{i}P(z_{i})})^2}.\label{eq10}
\end{eqnarray}
\end{widetext}
In this paper, $a$ is calculated by this statistic formula.
Substituting Eq. (\ref{eq02}) and Eq. (\ref{eq06}) into Eq.
(\ref{eq10}), we get,
\begin{eqnarray}
a&=&\frac{\langle{z\cdot{A(z)}}\rangle_{T}-\langle{z}\rangle_{T}\langle{A(z)}\rangle_{T}}
{\langle{z^{2}}\rangle_{T}-\langle{z}\rangle^{2}_{T}}\nonumber\\
&=&\frac{1}{\Delta{t}}[\frac{\langle{z(t)\cdot{z(t+\Delta{t})}}\rangle_{T}-\langle{z(t)}\rangle_{T}\langle{z(t+\Delta{t})}\rangle_{T}}
{\langle{z(t)^{2}}\rangle_{T}-\langle{z(t)}\rangle_{T}^{2}}-1].\nonumber\\
\label{eq11}
\end{eqnarray}
When $T$, compared to $\Delta{t}$, is sufficiently long, the
stationary assumption of financial time series is:
$\langle{z(t)}\rangle_{T}\approx{\langle{{z(t+\Delta{t})}}\rangle}_{T}$.
Then, compared Eq. (\ref{eq11}) with Eq. (\ref{eq08}), a simple
relation between $a$ and $C$ will be found,
\begin{eqnarray}
a(\Delta{t})\approx\frac{1}{\Delta{t}}[C(\Delta{t})-1]
\label{eq12}
\end{eqnarray}
which is valid for any value of $\Delta{t}$ because of no limit to
$\Delta{t}$ during the derivation. However, since the lack of
correlation for $\Delta{t}>1d$, only the case with $\Delta{t}=1d$
is analyzed. The statistical result $C_{\Delta{t}=1d}$ is compared
with the slope coefficient $a$, and the correlation  function
between $a$ and $C$ exhibited the good effectiveness of Eq.
(\ref{eq12}) [showed in Fig. \ref{fig6}]. The analytical results
indicate that the sign-cases, slope coefficient $a$, and
autocorrelation function $C_{\Delta{t}=1d}$ reflect the similar
properties of time series. Known that correlations observed in
financial time series show the incompleteness of the efficient
market hypothesis \cite{Boguna}, the three coefficients $C$, $a$,
and $\beta$ may probably indicate the degree of market efficiency.
From the same tendency of the three coefficients showed in Fig.
\ref{fig6} with $T=2000d$ (right), two conclusions will be arrived
at: (i) the market lost efficiency from $1961$ to $1976$
relatively, since the values is much larger than the remaining
$25$ years; (ii) the market tended to be more and more efficient
from $1968$ to $1999$ because of the decreasing trend of the
value.

From the preceding analysis of the equilibrium position $z_{0}$
and the slope coefficient $a$, the final form of restoring force
$A(z)$ will be got by substituting Eq. (\ref{eq07}) and Eq.
(\ref{eq12}) into Eq. (\ref{eq04}),
\begin{eqnarray}
A(z)=\frac{1}{\Delta{t}}{(C-1)(z-\langle{z}\rangle)}. \label{eq13}
\end{eqnarray}
This new form as a function of the traditional qualities, $C$ and
$\langle{z}\rangle$, is a more direct way to understand the
dynamical behavior of time series. To the financial data, the
information given by $A(z)$ mixes features of the macroscopical
properties together with the detail of the prices evolution: the
macroscopical trend of prices is presented by the equilibrium
position $z_0$, and the detail correlation between two neighboring
days is exhibited by the slope coefficient $a$.

Eq. (\ref{eq13}) can be informatively rewritten as
$A(z)=-(z-z_{0})/t_{0}$. $t_{0}$ is the characteristic relaxation
time, $t_{0}=-\Delta{t}/a$, and reflects the same properties of
the log-returns as the slope coefficient $a$ does, but $t_{0}$ is
more visualized. The maximum values of $t_{0}$, which were
calculated from the sliding windows of various length, are all
less than $2$ days, and the average values of $t_{0}$ with
$T=250d$ and $T=2000d$ are $\bar{t_{0}}|_{250d}=1.153d$ and
$\bar{t_{0}}|_{2000d}=1.157d$. Thus, the effect of the restoring
force $A(z)$ of one given day decays with the characteristic
relaxation time less than $2$ days.

\subsection{\label{sec:level6} The diffusion parameter $B(z)$}

The diffusion parameter $B(z)$, which usually has the form shown
in Eq. (\ref{eq05}), corresponds to a state-dependent linear
multiplicative noise term $B(z)dw$ in Eq. (\ref{eq01}). That is to
say, the Langevin description of the log-returns requires a linear
multiplicative noise term to describe the variability of the
log-returns, which can be interpreted as the variability of
log-returns increases with log-returns itself. Thus, we conjecture
that the heavy tailed probability densities are due to the form of
diffusion parameter $B(z)$ in our Langevin description.

\subsection{\label{sec:level7} Fitting error $\sigma$}

\begin{figure}
\centerline{\resizebox{9cm}{!}{\includegraphics{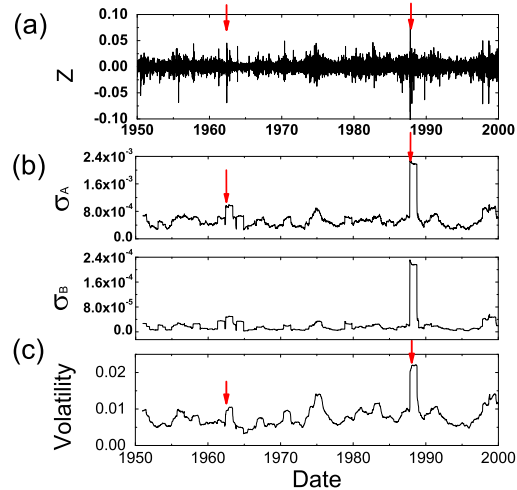}}}
\caption{(Color online) (a) Daily log-returns of S$\&$P $500$
index from $1950$ to $1999$. (b) The time-dependence of fitting
errors $\sigma_{A}$ and $\sigma_{B}$ with $T=250d$. (c) The time
averaged volatility of the same series with $T=250d$.}
\label{fig7}
\end{figure}

It is valuable to investigate the weighted fitting error of $A(z)$
and $B(z)$. Fig. \ref{fig7}(b) shows the time dependence of
$\sigma_{A}$ and $\sigma_{B}$ calculated from the sliding window
with $T=250d$ down the log-returns series. The time averaged
volatility of log-returns, which can measure the degree that the
market is liable to fluctuate, is calculated from the same $250d$
sliding window [see Fig. \ref{fig7}(c)]. Compared with the
original daily log-returns series [Fig. \ref{fig7}(a)], one can
easily find that the variation of $\sigma_{A}$ and $\sigma_{B}$
[Fig. \ref{fig7}(b)], together with the time averaged volatility
[Fig. \ref{fig7}(c)], show sudden jumps when very volatile periods
enter or leave the time window. For example, as pointed out by
arrows, the jumps at $T=9500d$ and $3110d$ are caused by the
crashes in May. $1962$ and on the `black Monday', $19$ Oct.
$1987$.

Based on the above analysis, one can conclude that the fitting
error, $\sigma_{A}$ and $\sigma_{B}$, sensitively respond to the
volatility of the financial markets, i.e., the information of
historical events in real market would influence the accuracy of
the fitting results presented in this paper.

\subsection{\label{sec:level8} The results of the log-return series $z(t)|_{\tau\neq{1d}}$}

In the way of our Langevin approach to the log-return series
$z(t)|_{\tau}$ with $\tau=2d,3d,4d,5d$ are also studied. It can be
concluded that, the forms of $A(z)$ and $B(z)$ [Eq. (\ref{eq04})
and Eq. (\ref{eq05})], the correlation function between the
equilibrium position $z_{0}$ and the average log-return
$\langle{z}\rangle$ [Eq. (\ref{eq07})], the correlation function
between slope $a$ and autocorrelation $C$ [Eq. (\ref{eq12})], and
the new form of restoring force $A(z)$ [Eq. (\ref{eq13})] are all
effective for these series. Table $1$ lists the results from the
log-return series of the S$\&$P $500$ index on the time scale
$\tau$ from $1d$ to $5d$.

\begin{table}
\caption{\label{tab:table2} The results from the log-return series
of the S$\&$P $500$ index on the time scale $\tau$ from $1d$ to
$5d$. The average slope $\bar{a}$, the average value of
autocorrelation $\bar{C}$, the average proportion of the
sign-sustained and sign-convert cases $\bar{\beta}$, the
characteristic relaxation time $t_{0}$ with $T=250d$ and $2000d$
are listed in this table.}
\begin{ruledtabular}
\begin{tabular}{cccccccccccc}
 \multicolumn{1}{c}{$\tau(d)$}&\multicolumn{2}{c}{$\bar{a}$}
&\multicolumn{2}{c}{$\bar{C}$}&\multicolumn{2}{c}{$\bar{\beta}$}
&\multicolumn{2}{c}{$\bar{t_0}(d)$}\\
&$250d$&$2000d$&$250d$&$2000d$&$250d$&$2000d$&$250d$&$2000d$\\\hline
%    a1     a2    C1   C2   beta1  beta2  t01  t02
1&-0.883&-0.874&0.115&0.125&1.242&1.223&1.153&1.157\\
2&-0.467&-0.462&0.533&0.539&2.273&2.252&2.178&2.187\\
3&-0.303&-0.299&0.697&0.702&3.205&3.137&3.395&3.414\\
4&-0.224&-0.222&0.776&0.779&4.136&4.059&4.416&4.633\\
5&-0.182&-0.177&0.818&0.825&4.986&4.886&5.759&5.822\\
\end{tabular}
\end{ruledtabular}
\end{table}

\section{\label{sec:level9} Summary and outlook}

In this paper, we present a coarse-grain time-dependent Langevin
description of the dynamics of stock prices, which is proved to be
effective by the results obtained from analyzing the S$\&$P $500$
index. The time dependence of drift parameter $A(z)$, which was
considered as the restoring force, was investigated by the simple
sliding windows algorithm. Significantly, while choosing the right
weighted factor (Eq. (\ref{eq06})) to approximate the statistical
results of $A(z)$, the linear approximations of $A(z)$ can reflect
both the macroscopical and the detail properties of the price
evolution, and the final form of the restoring force Eq.
(\ref{eq13}) can be achieved from analytical methods. The
macroscopical trend of price could be investigated from the
equilibrium position $z_0$, and the daily correlation in
log-return was exhibited by the flat slope coefficient $a$. The
mechanism of our model is discussed by analyzing the sign series
of log-returns. Therefore, from the restoring force $A(z)$ in
Langevin approach, one can get the properties of experimental data
or the properties of financial markets. Furthermore, it must be
pointed out that the random force $B(z)$ also plays an important
role in the dynamics of financial markets, which will be addressed
further in the future study.

\begin{acknowledgments}
We would like to thank Professor Hong Zhao for helpful
discussions. One of us (YC) acknowledges NSFC and Lanzhou
university for financial support.
\end{acknowledgments}

\newpage


\begin{references}
\bibitem{Bachelier}
L. Bachelier, Ann. Sci. $\acute{E}$cole Norm. Suppl. {\bf 3}, 21 (1900).

\bibitem{Pareto}
V. Pareto, {\it Manuel d'Economie Politique} (Marcel Giard, Paris, 1927).

\bibitem{Levy}
P. Le$\acute{v}$y, {\it Th$\acute{e}$orie de l'Addition des Variables
Al$\acute{e}$atoires} (Gauthier-Villars, Paris, 1937).

\bibitem{Mandelbrot}
B. B. Mandelbrot, J. Business {\bf 36}, 294 (1963).

\bibitem{Matengna}
R.N. Matengna and H.E. Stanley, {\it An Introduction to Econophysics} (Oxford University Press, New York,1971).

\bibitem{Zhang}
Y.-C. Zhang, Physica A {\bf 269}, 30 (1999).

\bibitem{Liu}
Y. Liu, P. Gopikrishnan, P. Cizeau, M. Meyer, C.-K. Peng, and H. E. Stanley, Phys. Rev. E {\bf 60}, 1390 (1999).

\bibitem{Gopikrishnan}
P. Gopikrishnan, V. Plerou, L.A. NunesAmaral, M. Meyer, and H. E. Stanley, Phys. Rev. E {\bf 60}, 5305 (1999).

\bibitem{Fama}
E. Fama, J. Finance {\bf 25}, 383 (1970).

\bibitem{LeBaron}
B. LeBaron, J. Business {\bf 65}, 199 (1992).

\bibitem{Boguna}
M. Boguna and J. Masoliver, Eur. Phys. J. B {\bf 40}, 347 (2004).

\bibitem{Ohira}
T. Ohira, N. Sazuka, K. Marumo, T. Shimizu, M. Takayasu, and H. Takayasu, Physica A {\bf 308}, 368 (2002).

\bibitem{Sazuka}
N. Sazuka, T. Ohira, K. Marumo, T. Shimizu, M. Takayasu, and H. Takayasu, Physica A {\bf 324}, 366 (2003).

\bibitem{Friedrich}
R. Friedrich, J. Peinke, and Ch. Renner, Phys. Rev. Lett. {\bf 84}, 5224 (2000).

\bibitem{Ivanova}
K. Ivanova, M. Ausloos, and H. Takayasu, cond-mat/0301268.

\bibitem{Sura}
P. Sura and J. Barsugli, Phys. Lett. A. {\bf 305}, 304 (2002).

\bibitem{Hanggi}
H. Risken, {\it The Fokker-Planck Equation}  (Springer-Verlag, Berlin, 1984).

\bibitem{Neter}
J. Neter, {\it Applied linear regression models}  (First edition
Richard D. IRWIN, INC, 1983).

\bibitem{Seber}
G.A.F. Seber, {\it Linear regression analysis}  (John Wiley $\&$ Sons, 1977).

\end{references}
\end{document}